\documentclass[twocolumn]{emulateapj}

\usepackage{epsfig}
\usepackage{longtable}
\usepackage{comment}
\usepackage{color}
\usepackage[normalem]{ulem}
\usepackage[colorlinks=true, pdfstartview=FitV, linkcolor=blue,citecolor=blue, urlcolor=blue]{hyperref}

\newcommand{\cxc}{{\em Chandra}}

\shorttitle{Chandra observations of the X-ray halo hosting 4C+37.11}
\shortauthors{Andrade-Santos et al.}

\begin{document}


\title{Binary black holes, gas sloshing, and cold fronts \\
in the X-ray halo hosting 4C+37.11}


\author{
Felipe Andrade-Santos$^1$, 
\'Akos Bogd\'an$^1$,
Roger W. Romani$^2$,
William R. Forman$^1$,
Christine Jones$^1$,\\
Stephen S. Murray$^1$,
Greg B. Taylor$^3$, and
Robert T. Zavala$^4$}
\affil{$^1$Harvard-Smithsonian Center for Astrophysics, 60 Garden Street, Cambridge, MA 02138, USA \\
$^2$Department of Physics, Stanford University, Stanford, CA 94305-4060, USA \\
$^3$Department of Physics and Astronomy, University of New Mexico, Albuquerque, NM 87131, USA \\
$^4$US Naval Observatory, Flagstaff Station, 10391 W. Naval Observatory Rd, Flagstaff, AZ 86001, USA
}




\begin{abstract}

We analyzed deep \textit{Chandra} ACIS-I exposures of the
cluster-scale X-ray halo surrounding the radio source 4C+37.11. This remarkable system hosts the closest resolved pair of super-massive black
hole and an exceptionally luminous elliptical galaxy, the likely product of a series of past mergers.
We characterize the halo with $r_{500} = 0.95$ Mpc, $M_{500} = (2.5 \pm 0.2) \times
10^{14} \ M_{\rm{\odot}}$, $ kT = 4.6\pm 0.2$ keV, and a gas mass of
$M_{\rm g,500} = (2.2 \pm 0.1) \times 10^{13}
M_\odot$. The gas mass fraction within $r_{500}$ is 
$f_{\rm g} = 0.09 \pm 0.01$. The entropy profile shows large non-gravitational
heating in the central regions. We see several surface brightness
jumps, associated with substantial temperature
and density changes, but approximate pressure equilibrium, implying that these are sloshing structures 
driven by a recent merger. A residual intensity image shows core spiral structure closely matching
that seen for the Perseus cluster, although at $z=0.055$ the spiral
pattern is less distinct. We infer the most recent merger occurred
$1-2$ Gyr ago and that the event that brought the
two observed super-massive black holes to the system core is even
older. Under that interpretation, this black hole binary pair has, unusually,
remained at pc-scale separation for more than 2 Gyr.

\end{abstract}



\section{Introduction}

Supermassive Black Hole Binaries (SMBHBs) likely form through
galaxy mergers \citep{1980Begelman} and show a wide range of
separations \citep{1985Owen,2003Komossa}.  Recently \citet{2015Yan}
suggested that a supermassive black hole binary with subparsec
separation could explain the optical-UV spectrum of the quasar Mrk 231.
Currently the SMBHB with the smallest directly measured projected
separation (7.3 pc) lies in the core of the radio galaxy 4C+37.11
\citep{2006Rodriguez}. Through VLBA observations, \citet{2004Maness} discovered the 
two flat spectrum variable components in this system. 
Through additional VLBA observations with higher angular resolution, 
\citep{2006Rodriguez} concluded that the two radio sources were 
the nuclei of a supermassive black hole binary system.  Since the 
massive galaxies whose merger produces SMBHBs are often at the centers 
of galaxy groups or clusters, one can also observe the effects of the 
merger on larger scales.  In particular the thermodynamic signatures 
of the merger in the cluster gas in the form of gas sloshing can be 
observed for billions of years following the merger \citep[e.g.,][]{2006Ascasibar}. 
Gas sloshing has been extensively studied both through observations
\citep[e.g.,][]{2003Churazov,2010Lagana,2013Paterno-Mahler} and simulations \citep[e.g.,][]{2006Ascasibar,2010ZuHone,2012Roediger}.  

We present the large-scale properties of the cluster-scale X-ray halo
hosting the radio galaxy 4C+37.11. 4C+37.11
is a remarkable system for two major reasons. First, it hosts two compact radio nuclei, 
resolved by Very Long Baseline Array (VLBA) observations to have a 7.3 pc 
projected separation, one of which currently 
powers relativistic jets in a Compact Symmetric Object outflow (CSO)
\citep{2004Maness,2006Rodriguez}. 
The optical host of the SMBHB is a relatively isolated, extremely
luminous elliptical galaxy with  $M_{K_s}$ = -27.0, 
which corresponds to a stellar mass of $M_\odot = 9.5 \times 10^{11}
M_\odot$.
Given the environment of the system, the source 
may be a fossil group \citep{2014Romani} with an unmerged 
pair of nuclear black holes. Second, X-ray observations of 4C+37.11 show
that the binary black hole is embedded in a bright, extended X-ray
halo. Specifically, the X-ray luminosity within $R_{2500} = 400$ kpc   is
$L_{\rm X} \sim ~10^{44} \rm ~ erg~s^{-1}$.
These characteristics show that 4C+37.11 resides in galaxy cluster-sized dark matter halo. 
Previous {\em Chandra} observations revealed that the hot gas shows edge
structures \citep{2014Romani}, hinting at a past merger, 
which may have led to the formation of the observed SMBHB. However, 
the previous {\em Chandra} observation did not have sufficiently high 
signal-to-noise  ratio to constrain the origin of the edges. 
The goal of this paper is to utilize our deep {\em Chandra} X-ray
observations to probe the origin of the edge structures, 
and constrain the characteristics of the merger that led to the
formation of a binary black hole.

The paper is structured as
follows. In Section 2 we describe the data
reduction. In Section 3, overall properties of the halo hosting
4C+37.11 are described. In Section 4 we present evidence of a sloshing 
feature in 4C+37.11. The small scale structure of the cluster is 
investigated in Section 5. We discuss the implications of our results 
in Section 6, and we conclude in Section 7. 

Assuming a standard $\Lambda$CDM cosmology with $H_0= ~ 70 \rm ~km~s^{-1}~Mpc^{-1}$,
$\Omega_{\rm M}=0.3$, and $\Omega_{\rm \Lambda}=0.7$, the observed redshift of 
4C+37.11, $z=0.055$, implies a
linear scale of $1.07 \rm ~kpc~arcsec^{-1}$ and a luminosity distance of 246 Mpc.   

\section{Data Reduction}
\label{sec:chandra} 

4C+37.11 was observed on April 4, 2011, and November 6, 2013 with the \cxc ~Observatory
(both observations, the 10 ks ObsId \dataset[ADS/Sa.CXO#obs/12704]{12704} (PI: Murray) and the 95 ks
ObsId \dataset[ADS/Sa.CXO#obs/16120]{16120} (PI: Romani) were performed with the ACIS-I array, 
thereby offering a large field-of-view). 

The data reduction followed the process described in
\citet{2005Vik}. We applied the calibration files \texttt{CALDB 4.6.2}. 
The data reduction includes corrections for the time dependence of the charge
transfer inefficiency and gain, and also a check for periods
of high background (none were found). Standard blank sky background files
and readout artifacts were subtracted.

\section{Overall characteristics of the cluster}\label{sec:gas_properties}

\subsection{Emission measure profile}

We refer to \citet{2006Vik} for a detailed description of
the procedures used to compute the emission measure profile.
We outline here only the main aspects of the method. 

First we detected compact sources in the 0.7--2.0 keV or 2.0--7.0 keV bands 
and then masked these from the spectral and spatial analyses. We then measured
the surface brightness profiles in the 0.7--2.0 keV energy band, 
which maximizes the signal to noise ratio in
{\em Chandra} data. The readout artifacts and blank-field background \citep[see section 2.3.3 of][]{2006Vik} 
are subtracted from the X-ray images, and the result is exposure-corrected using exposure maps
(computed assuming an absorbed optically-thin thermal plasma with $kT = 5.0$
keV, abundance = 0.3 solar, plus the Galactic column density)
that include corrections for bad pixels and CCD gaps, but do not 
take into account spatial variations of the effective area. 
Then, we subtract any small uniform component corresponding to 
soft X-ray foreground adjustments that may be required.

Following these steps, we extract the surface brightness profiles in narrow 
concentric annuli ($r_{\rm out}/r_{\rm in} = 1.05$) centered on the
X-ray halo peak and compute the {\em Chandra} area-averaged effective area for each
annulus \citep[see][for details on calculating the effective area]{2005Vik}. 
Using the observed projected temperature, effective area, and 
metallicity as a function of radius, we then convert the {\em Chandra} count 
rate in the 0.7--2.0 keV band into the emission integral, 
${\rm EI} =  \int n_{\rm e} n_{\rm p} dV$, within each cylindrical
shell. To compute the emission measure and temperature profiles we assume spherical 
symmetry, although the X-ray morphology of 4C+37.11 exhibits a mildly elliptical shape. 
The spherical assumption is expected to introduce only negligible (less than 3\%) 
deviation \citep{2003Piffaretti}.

We then fit the emission measure profile assuming the gas density
profile follows \citet{2006Vik}: 

\begin{eqnarray}
n_{\rm e}n_{\rm p} &=& n_0^2
\frac{ (r/r_{\rm c})^{-\alpha}}{(1+r^2/r_{\rm c}^2)^{3\beta-\alpha/2}}
\frac{1}{(1+r^\gamma/r_{\rm s}^\gamma)^{\epsilon/\gamma}}+ \nonumber \\
&&\frac{n^2_{02}}{(1+r^2/r_{\rm c2}^2)^{3\beta_2}}.\label{eq:nenp}
\end{eqnarray} 

This relation is based on a classic $\beta$-model, however it is modified to account
for the power-law type cusp and the steeper emission measure slope at large radii. In addition 
a second $\beta$-model is included, giving extra freedom to characterize the cluster core. 
For further details on this equation we refer to \citet{2006Vik}. The
relation between the electron number density and gas mass density is given by $\rho_{\rm g} =
\mu_{\rm e} n_{\rm e} m_{\rm a}$, where $m_{\rm a}$ is the atomic mass
unit and $\mu_{\rm e}$ is the mean
molecular weight per electron. For a typical metallicity of
0.3 $Z_\odot$, the reference values from \citet{1989AndersGrevesse}
yields $\mu_{\rm e} = 1.17058$ and $n_{\rm e}/n_{\rm p} = 1.1995$.

The best fit parameters of Equation \ref{eq:nenp} are presented in Table \ref{tab:emm_prof}.

\begin{figure}[!]
  \begin{center}
    \leavevmode
      \epsfxsize=8.75cm\epsfbox{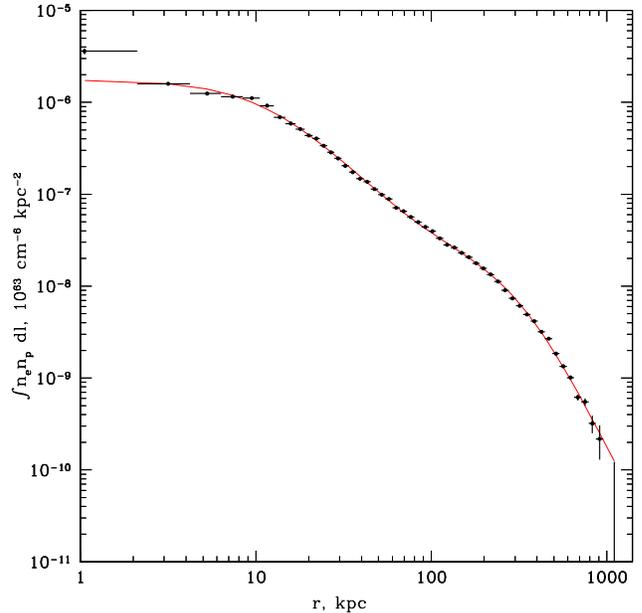}
      \caption{Observed projected emissivity profile for 4C+37.11. The
      solid line shows the emission measure integral of the best fit
      to the emissivity profile given by Equation (\ref{eq:nenp}). The emission from the nucleus is 
      excluded from the fit, since it exhibits an enhanced emission due to the AGN.}\label{fig:emm_prof}
  \end{center}
\end{figure}

\begin{deluxetable*}{cccccccccc}
\tablecaption{Parameters for the emission measure profile (Equation \ref{eq:nenp})} 
\tablewidth{0pt} 
\tablehead{ 
\colhead{$n_0$} &
\colhead{$r_{\rm c}$} & 
\colhead{$r_{\rm s}$} &
\colhead{$\alpha$} &
\colhead{$\beta$} &
\colhead{$\gamma$} & 
\colhead{$\epsilon$} &
\colhead{$n_{02}$} &
\colhead{$r_{\rm c2}$} & 
\colhead{$\beta_2$} \\
\colhead{($\rm 10^{-2} ~ cm^{-3}$)} &
\colhead{(kpc)} &
\colhead{(kpc)} &
\colhead{} &
\colhead{} &
\colhead{} &
\colhead{} &
\colhead{($\rm 10^{-3} ~ cm^{-3}$)} &
\colhead{(kpc)} &
\colhead{} 
}
\startdata 
$5.28 \pm 0.62$ & $11.7 \pm 1.4$ & $46.4 \pm 3.8$ & $0.02 \pm 0.11$ & $0.359 \pm 0.027$ & $0.94 \pm 0.19$ & $1.66 \pm 0.37$ & $1.425 \pm 0.083$ & $331 \pm 18$ & $0.935 \pm 0.039$
\enddata
\tablecomments{Columns list best fit values for the parameters
given by Equation \ref{eq:nenp}.} 
\label{tab:emm_prof}
\end{deluxetable*}

\subsection{Gas Temperature Radial Profiles}\label{sec:temp}

\begin{figure}[!]
  \begin{center}
    \leavevmode
      \epsfxsize=8.75cm\epsfbox{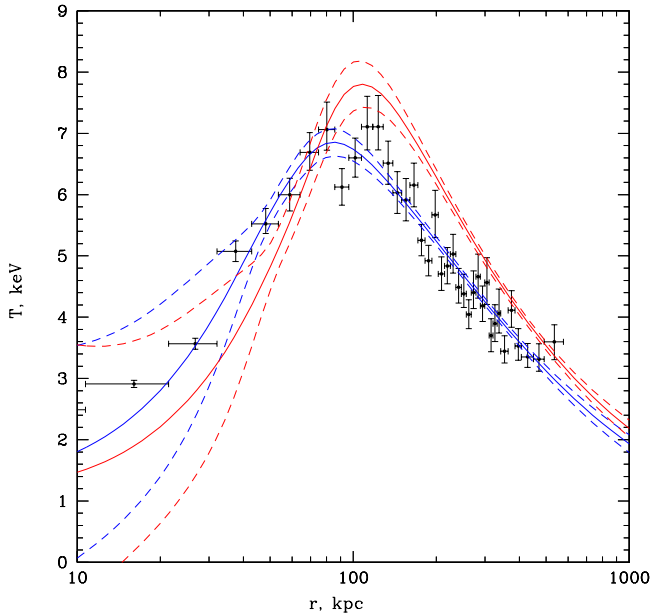}
      \caption{Azimuthally averaged, radial temperature profile. Observed
  projected temperatures are shown by points with error
  bars. The 3D model and its projected
effective temperatures (the latter to be compared with the data) are shown by the
red and blue curves, respectively. Dashed lines show the 1$\sigma$ uncertainty ranges.}\label{fig:temp_prof}
  \end{center}
\end{figure}

Most clusters present a temperature profile that has
a broad peak within 0.1--0.2 $r_{200}$\footnote{$r_{200}$ and
$r_{500}$ are used to define a radius at the over-density of 200 and
500 times the critical density of the Universe at the cluster
redshift, respectively.}. 
\citep{2006Vik} presents a 3D temperature
profile that describes these general
features. At large radii, the temperature profile
can be fairly represented as a broken power law with a transition
region:
\begin{equation}\label{eq:tprof:main}
  T(r) =\frac{(r/r_t)^{-a}}{(1+(r/r_t)^b)^{c/b}}. 
\end{equation}

At small radii, the temperature profile can be described as
\begin{equation}\label{eq:tprof:cool}
  T_{\text{cool}}(r) = (x+T_{\text{min}}/T_0)/(x+1),
\end{equation}
where $x=(r/r_{\text{cool}})^{a_{\text{cool}}}$. The final analytical expression for the 3D temperature
profile is,
\begin{equation}\label{eq:tprof}
  T_{\mathrm{3D}}(r) =  T_0\times T_{\text{cool}}(r)\times T(r). 
\end{equation}

This temperature model has significant functional freedom (8 parameters) and can
adequately describe almost any smooth temperature distribution. 
Thus, we use this model, from \citet{2006Vik}, to describe the temperature distribution of the hot gas in 4C+37.11. 

To estimate the uncertainties
in the best values for the parameters of this analytical model, we
performed Monte-Carlo simulations.
This model for $T_{\mathrm{3D}}(r)$ (Equation \ref{eq:tprof}) allows
very steep temperature gradients. In some Monte-Carlo realizations,
such profiles are mathematically consistent with the observed projected
temperatures, however, large
values of temperature gradients often lead to unphysical mass
estimates, such as profiles with negative dark matter density at some
radii. We solved this issue by accepting only Monte-Carlo realizations in which the
best-fit temperature profile leads to $\rho_{\text{tot}}>\rho_{\text{gas}}$ in the radial
range $r \leq 1.5 r_{500}$, where $\rho_{\text{tot}} = \rho_{\rm gas} + \rho_{\rm dark~ matter}$.
Also, in the same radial range, we
verified that the temperature profiles are all convectively stable, i.e. $d\ln  T/d\ln  \rho_g < 2/3$.

To construct the temperature profile, we extracted spectra from 38 annuli
in the radial range from 0 to
$\sim 600$ kpc 
and fit them with an absorbed \texttt{APEC} model.
For the fitting we fixed the column density at $N_{\rm H} = (8.2 \pm
0.4) \times 10^{21}\rm ~cm^{−2}$. This value was obtained by fitting
the central 500 kpc region of the cluster with an absorbed \texttt{APEC} model
(with abundance fixed at $A$ = 0.3 Solar)
leaving the column density as a free parameter. This fit resulted in a
best-fit temperature of $kT = 4.6 \pm 0.2$ keV and $N_{\rm H} = (8.2 \pm 0.4) \times 10^{21}\rm ~cm^{−2}$. 
This gives an equivalent $A_{\rm V} = N_{\rm H}/(2.2 \times 10^{21} \rm ~ cm^{2}) =
3.7$, in excellent agreement with the full Galactic extinction
estimated from the reddening maps \citep{1998Schlegel,2015Green}, 
suggesting that the intrinsic absorption is small.
We then followed the procedures described above to obtain the 2D and
3D temperature profiles. 
The measured 2D (black data points), fitted 2D (blue solid line), and
3D (red solid line) temperature profiles are presented
in Figure \ref{fig:temp_prof}. The 2D temperature
profile was computed by projecting the 3D temperature weighted by gas density squared
using the spectroscopic-like temperature \citep[][provides a formula for the
temperature which matches the spectroscopically
measured temperature within a few percent]{2004Mazzotta}:
\begin{equation}
T_{\rm 2D}=T_{\rm spec} \equiv \frac{\int \rho_{\rm g}^2 T_{\rm
    3D}^{1/4} dz}{\int \rho_{\rm g}^2 T_{\rm 3D}^{-3/4} dz}
\label{eq:tspec}
\end{equation}

The best fit parameters of equations \ref{eq:tprof:main} and
\ref{eq:tprof:cool} are presented in Table \ref{tab:temp_prof}.

\begin{deluxetable*}{cccccccc}
\tablecaption{Parameters for the Temperature Profile (Equations \ref{eq:tprof:main} and \ref{eq:tprof:cool})} 
\tablewidth{0pt} 
\tablehead{ 
\colhead{$T_0$} &
\colhead{$T_{\rm min}$} & 
\colhead{$r_{\rm t}$} &
\colhead{$r_{\rm cool}$} &
\colhead{$a_{\rm cool}$} &
\colhead{$a$} & 
\colhead{$b$} &
\colhead{$c$} \\
\colhead{(keV)} &
\colhead{(keV)} &
\colhead{(kpc)} &
\colhead{(kpc)} &
\colhead{} &
\colhead{} &
\colhead{} &
\colhead{} 
}
\startdata 
$37 \pm 37$ & $14.2 \pm 7.7$ & $75 \pm 58$ & $97 \pm 100$ & $4.8 \pm
3.4$ & $-2.1 \pm 3.2$ & $3.1 \pm 1.9$ & $2.9 \pm 2.4$
\enddata
\tablecomments{Columns list best fit values for the parameters
given by Equations \ref{eq:tprof:main} and \ref{eq:tprof:cool}.} 
\label{tab:temp_prof}
\end{deluxetable*}

\subsection{Total and Gas Masses}\label{sec:total_mass}

Assuming hydrostatic equilibrium and using the three-dimensional analytical expressions for the temperature
and gas density profiles, one can compute the total enclosed mass
within a radius $r$ with the following equation (e.g., Sarazin
1988):
\begin{eqnarray}\label{e_m_r}
M(r) &=& \frac{-kTr}{\mu m_H G} 
\left( 
\frac{d\ln \rho_{\rm g}}{d\ln r}+
\frac{d\ln T}{d\ln r}
\right) \nonumber \\ 
&=& -3.67 \times 10^{13} ~ M_\odot kT r 
\left( 
\frac{d\ln \rho_{\rm g}}{d\ln r}+
\frac{d\ln T}{d\ln r}
\right), 
\end{eqnarray}
where $T$ is the temperature in units of K, $k$ is the Boltzmann
constant, and $r$ is in units of Mpc. The normalization
corresponds to  $\mu = 0.6107$, which was computed for an abundance of 0.3 $Z_\odot$.  

\begin{figure}[!]
  \begin{center}
    \leavevmode
      \epsfxsize=8.75cm\epsfbox{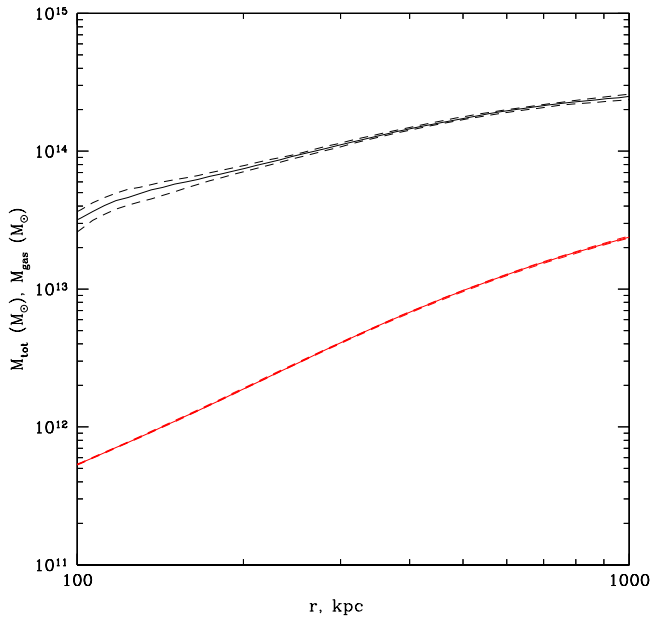}
      \caption{Computed mass profiles. Black and red lines correspond
        to total and gas mass profiles, respectively. Dashed
      lines correspond to 68\% confidence range.}\label{fig:mass_prof}
  \end{center}
\end{figure}

Using Equation \ref{e_m_r}, $r_{500}$ is directly computed by
solving 
\begin{eqnarray}
M(r_{500})=500 \rho_c (4\pi/3) r_{500}^{3},\label{m500_def}
\end{eqnarray}
where $\rho_c$ is the critical density of the Universe at the cluster
redshift. Using Equations \ref{e_m_r} and  \ref{m500_def}, we
obtain $r_{500} = 945_{-22}^{+21}$ kpc 
and a corresponding hydrostatic mass of 
$M_{\rm 500,hyd} = (2.53 \pm 0.17) \times 10^{14} M_\odot$ (Figure \ref{fig:mass_prof}). 

Using the best fit parameters for the density profile (see Table
\ref{tab:emm_prof}), we compute a gas mass within $r_{500}$ of $M_{\rm
  g,500} = (2.24 \pm 0.06) \times 10^{13}
M_\odot$. The gas mass fraction within $r_{500}$ is 
$f_{\rm g} = 0.09 \pm 0.01$ and is in agreement with the expected value
\citep[see Fig. 21 from][]{2006Vik} for clusters with $kT \sim$ 4.5 keV
($f_{\rm g} \sim$ 0.09--0.10). The results are summarized in Table \ref{tab:phys_prop}. 

\begin{deluxetable*}{cccccc}[h!]
\tablecaption{Physical Properties Derived Assuming Hydrostatic Equilibrium} 
\tablewidth{0pt} 
\tablehead{ 
\colhead{$r_{\rm 500}$} &
\colhead{$M_{\rm g,500}$} &
\colhead{$M_{\rm 500}$} &
\colhead{$f_{\rm g}$} &
\colhead{$kT$} &
\colhead{$L_{\rm X}$} \\
\colhead{(kpc)} &
\colhead{($10^{13} ~ M_\odot$)} &
\colhead{($10^{14} ~ M_\odot$)} &
\colhead{} &
\colhead{(keV)} &
\colhead{($10^{44} \rm ~erg~s^{-1}$)}
} 
\startdata 
$945_{-22}^{+21}$ & $2.24 \pm 0.06$ & $2.53 \pm
0.17$ & $0.09 \pm 0.01$ & $4.6 \pm 0.2$ & $1.10 \pm 0.01$
\enddata
\tablecomments{Columns list the cluster $r_{500}$, gas mass, total
  mass derived from the hydrostatic equilibrium equation (Equation (\ref{e_m_r})), gas
fraction, spectroscopic-like temperature, and bolometric X-ray luminosity within $r_{500}$.}
\label{tab:phys_prop} 
\end{deluxetable*}

\subsection{Entropy Profiles}

\begin{figure*}[htb!]
\centerline{\includegraphics[width=0.5\textwidth]{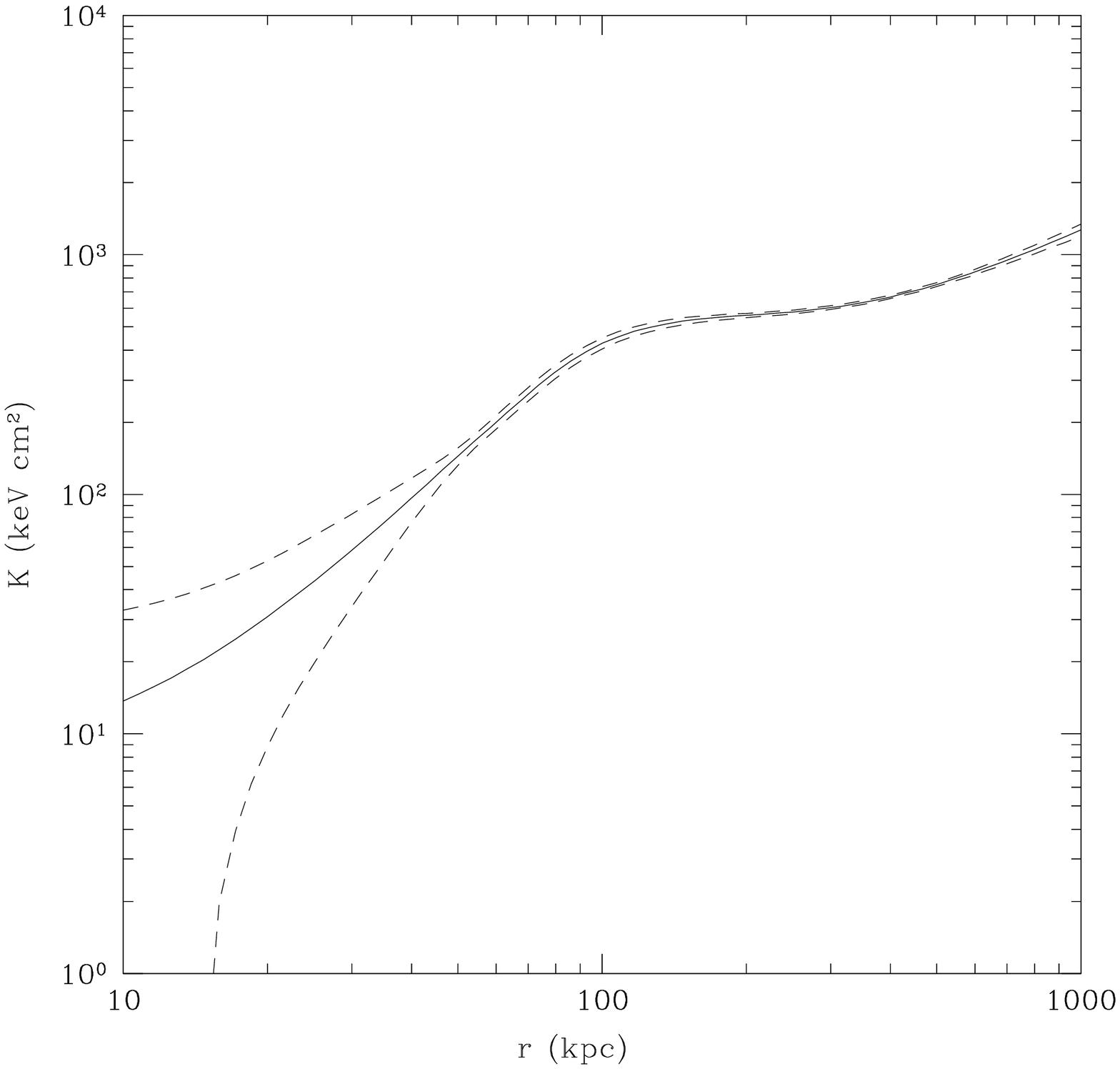}
\includegraphics[width=0.5\textwidth]{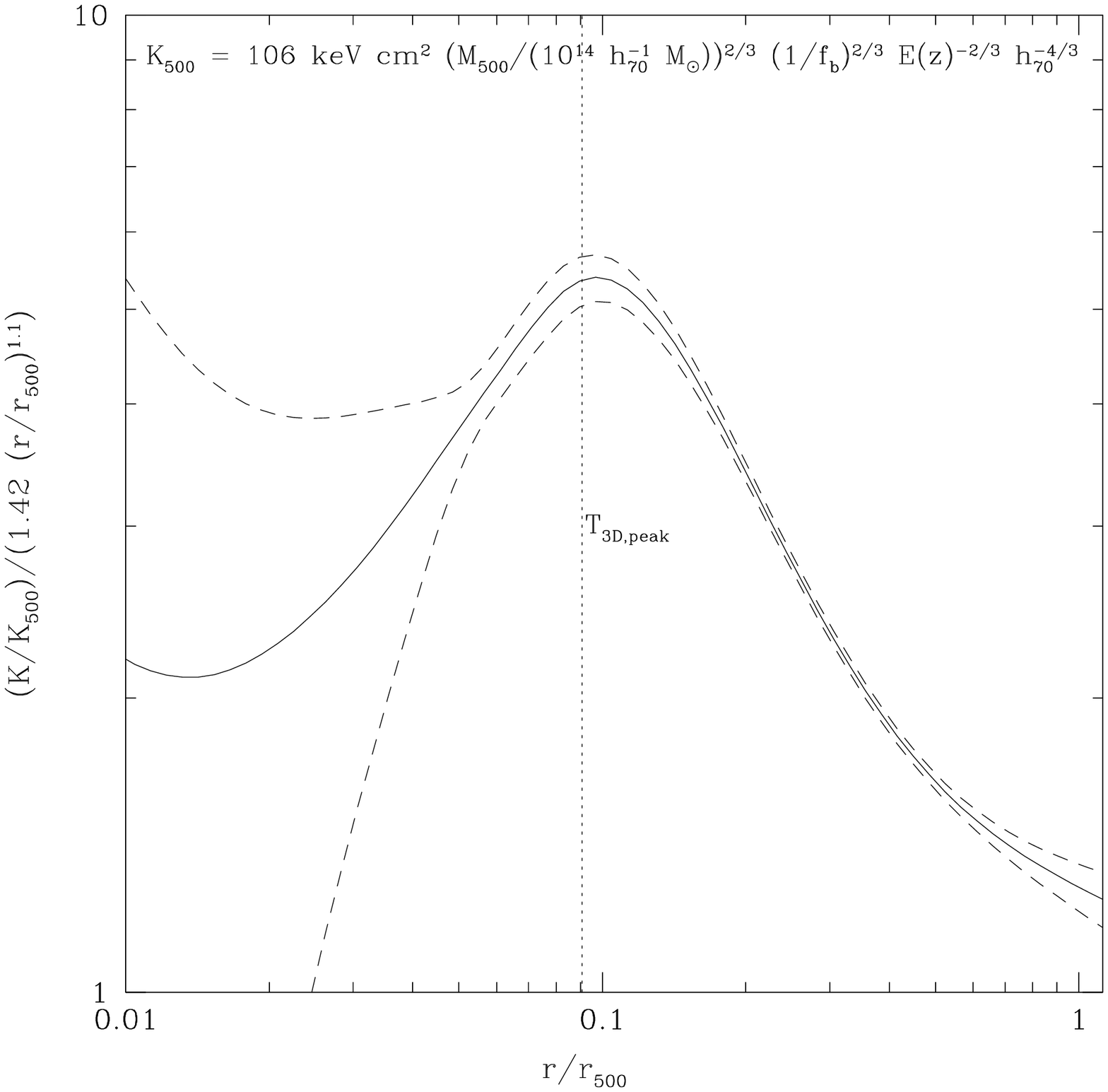}}
\caption{
\small{
Left: Entropy profile of the halo hosting 4C+37.11. Right:
Dimensionless entropy profile of 4C+37.11. We see
that the entropy index is higher at all radii than the scaling relation
$K/K_{500} = 1.42 (r/r_{500})^{1.1}$, which suggests
that non-gravitational processes are playing a significant role in the
thermodynamics of the ICM even at large distances from the center of
the cluster. Dashed lines correspond to 68\% confidence ranges. 
}
}
\label{fig:entropy_prof}
\end{figure*}

The intracluster gas entropy index is defined as
\begin{eqnarray}
K = \frac{kT}{n_{\rm e}^{2/3}},
\end{eqnarray} 
where $n_{\rm e}$ is the electron density, $T$ is the gas temperature,
and $k$ is the Boltzmann constant. The entropy index is directly
related to the thermodynamic history of the ICM. The entropy index increases when
heat energy is deposited non-adiabatically into the ICM, and decreases when
radiative cooling carries heat energy away \citep{2005Voit}. To understand the thermodynamic 
history of the halo hosting 4C+37.11, we
computed entropy profiles, which are presented in Figure
\ref{fig:entropy_prof}. 

Departure from the self-similar scaling relation $K/K_{500} = 1.42 (r/r_{500})^{1.1}$  \citep{2010Pratt} 
is suggestive of non-gravitational processes, where $K_{500}$ is
computed by:
\begin{eqnarray}
K_{500} &=& 106 ~ {\rm keV cm^2} \left( \frac{M_{500}}{10^{14} h_{70}^{-1}
M_\odot} \right)^{2/3} \left( \frac{1}{f_{\rm b}} \right)^{2/3}
\\ \nonumber
&\times& E(z)^{-2/3} h_{70}^{-4/3}, 
\end{eqnarray}
where $f_{\rm b} = 0.15$ is the baryon fraction and $E(z)=(\Omega_{\rm M}(1+z)^3+\Omega_{\rm k}(1+z)^2+\Omega_\Lambda)^{1/2}$. 
Figure \ref{fig:entropy_prof} shows the
dimensionless entropy profile of the galaxy halo hosting 4C+37.11.  We see
that the entropy index is higher, at all radii than the scaling relation
$K/K_{500} = 1.42 (r/r_{500})^{1.1}$, which suggests
that non-gravitational processes are playing a significant role in the
thermodynamics of the ICM, even at large distances from the center of
the cluster. We see substantial energy injection in the center of the cluster extending several arcmin,
to $\sim0.3r_{\rm{500}}$. Note that this is well outside of the $\sim10\arcsec$ radius of the radio lobes currently
being inflated by the nucleus, so it is not the direct product of the current AGN
activity (Section \ref{sec:small}). This entropy excess is suppressed by core cooling within $\sim100$ kpc,
which returns the entropy close to the \citet{2010Pratt} scaling set by $K_{500} $ at large radius.

\begin{figure*}[!]
  \begin{center}
    \leavevmode
      \epsfxsize=17.5cm\epsfbox{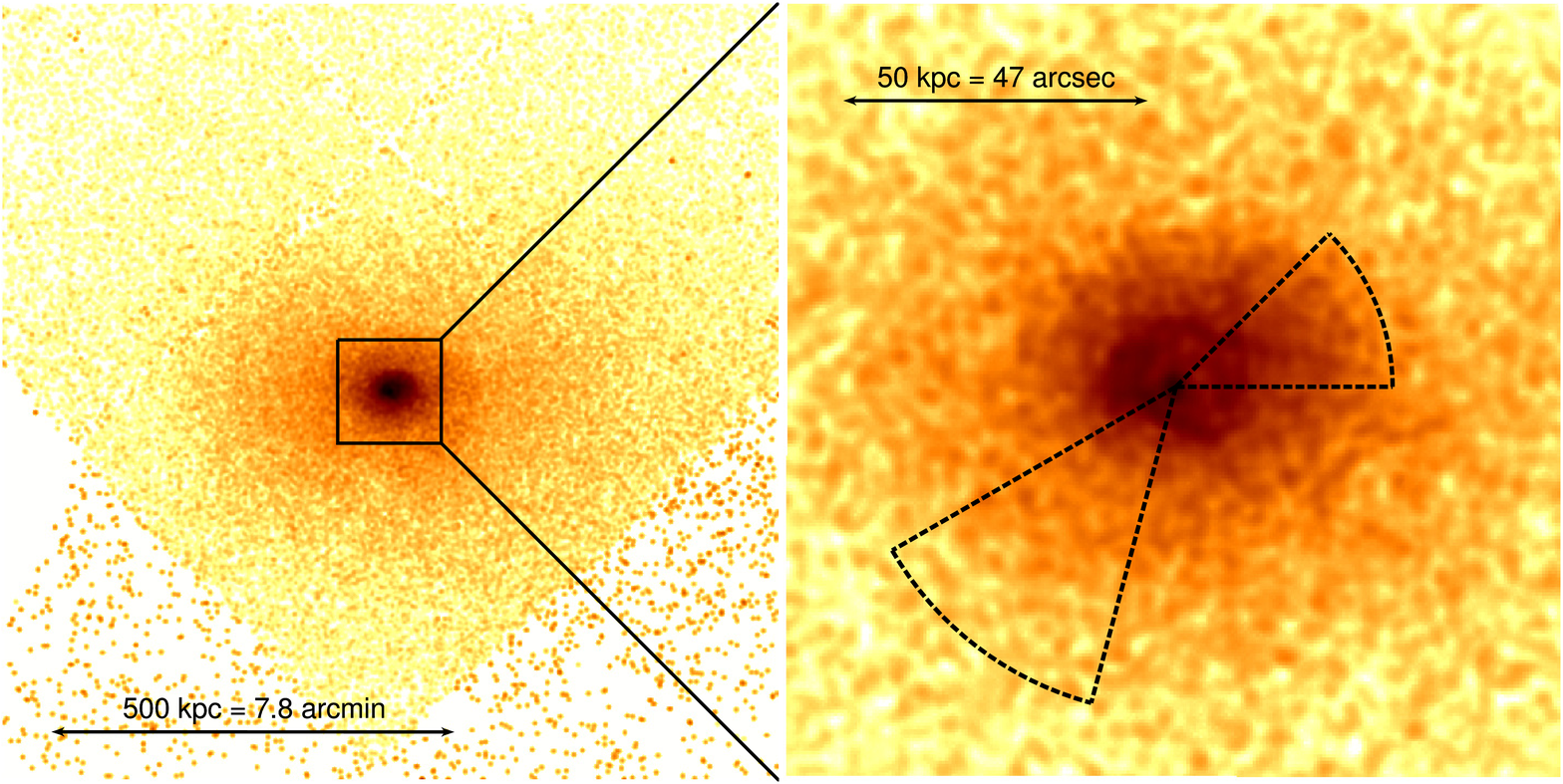}
      \caption{The left panel shows the $0.5-2$ keV band merged \textit{Chandra} image 
      of a $15\arcmin \times 15\arcmin$ ($960 \times 960$ kpc) region of 4C+37.11. To 
      highlight the large-scale features in the hot gas distribution, the image was 
      smoothed with a Gaussian kernel of two-pixel size. Bright point sources have been 
      excluded and their location has been filled with the local background level. The 
      box shows a $2\arcmin \times 2\arcmin$ ($128 \times 128$ kpc) region, which is 
      depicted in the right panel. To make the small-scale features more pronounced, 
      this image was smoothed with a Gaussian kernel of one-pixel size. The over plotted 
      regions  show the location and position angle of the wedge regions that were used 
      to extract surface brightness and density profiles. The outer shell of the regions 
      marks the position of the surface brightness edges, which are detected with the 
      surface brightness profiles in Figure \ref{fig:sb_profiles}. Note that the positions 
      of the surface brightness jumps are azimuthally asymmetric.}     \label{fig:image}
  \end{center}
\end{figure*}

\section{Results}
\subsection{Images}
\label{sec:images}
The left panel of Figure \ref{fig:image} shows the merged, flat-fielded 
(vignetting and exposure corrected), and background subtracted $0.5-2$ keV band 
\textit{Chandra} ACIS-I image of 4C+37.11. The image, depicting the inner 
$960 \times 960$ kpc region of the cluster, reveals the presence of large scale 
diffuse emission, which originates from optically-thin thermal plasma with $kT \sim 3-7$ 
keV temperature (Section \ref{sec:gas_properties}).

The distribution of the hot X-ray emitting gas reveals a complex
morphology, indicating an active merger history. In particular, the gas distribution 
is not symmetric, but is elongated in the east-west direction. In addition, the image 
shows the presence of sharp surface brightness edges in the central regions of the 
cluster. To further investigate the surface brightness features in the central regions 
of the cluster, we show a zoomed-in version of the large-scale image in the right 
panel of Figure  \ref{fig:image}. This image depicts the central $128 \times 128$ kpc 
region of  4C+37.11, and confirms our findings. Indeed, the hot gas distribution 
exhibits asymmetry at scales of 10 kpc, and hints at the presence of sharp surface 
brightness edges.

The above features are characteristic signatures of a merger, which has likely 
perturbed the hot gas distribution. To explore the nature of these features, and hence, 
constrain the merger history of the cluster, we derive surface brightness, density, and 
temperature profiles, which are discussed in the following sections.

\subsection{Profiles}
\label{sec:profiles}
In galaxy clusters, sharp surface brightness edges may be caused by
three phenomena: cold fronts induced by mergers, sloshing of the gas
in the central regions of clusters induced by minor mergers, and shocks associated with mergers
and supersonic inflation of radio lobes. To probe the origin of the
surface brightness edges in 4C+37.11, we build surface brightness,
density, and temperature profiles, and derive the pressure jump across the
edges. If the edges originate from a major sub-cluster merger, a
temperature and pressure jump is expected across the surface
brightness discontinuity. If, however, the edge is due to sloshing, we
expect to detect a change in the temperature and density across the
edge, but the pressure should remain at or near equilibrium. 

\begin{figure*}[!]
  \begin{center}
    \leavevmode
      \epsfxsize=8.2cm\epsfbox{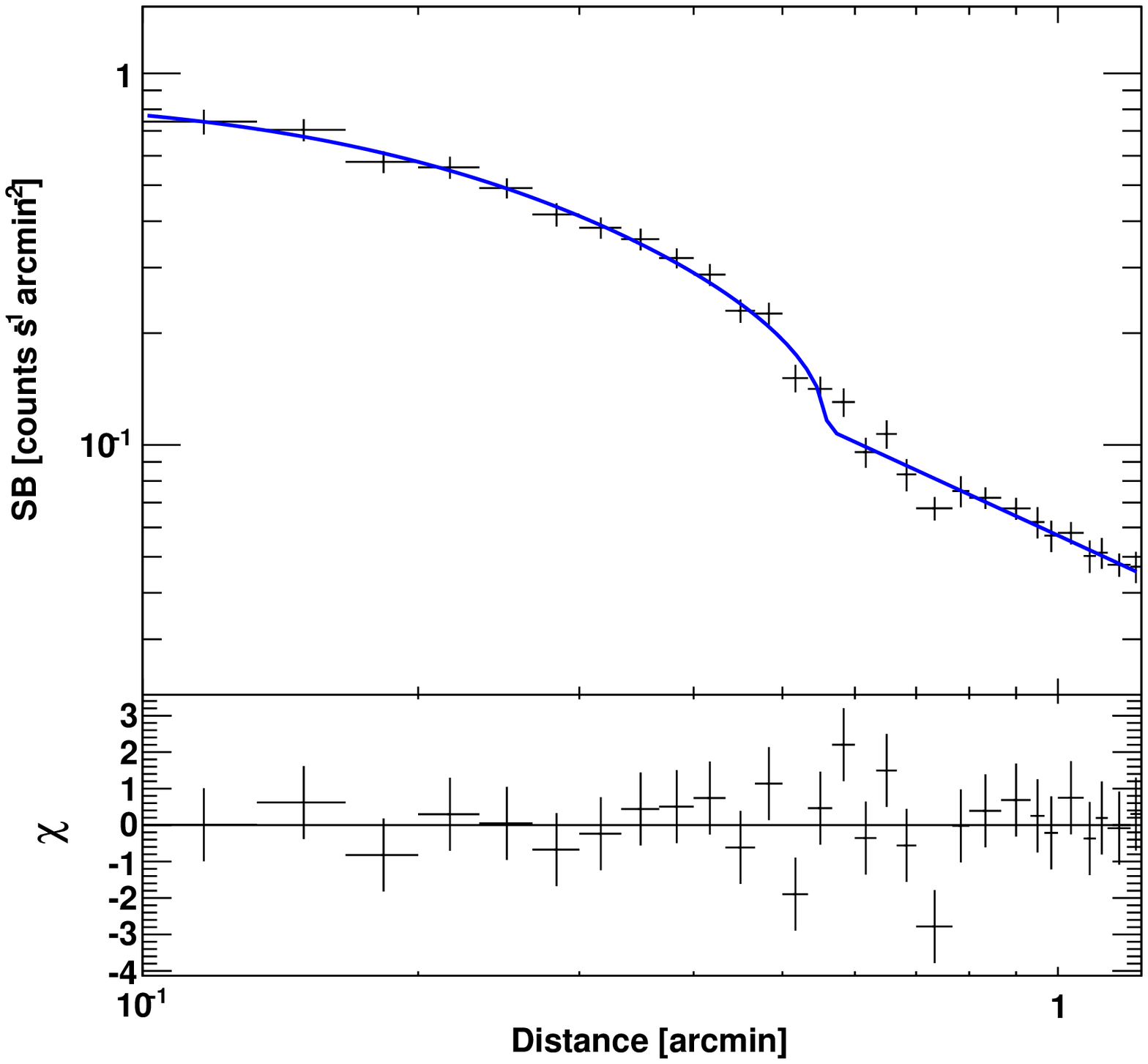}
      \hspace{0.5cm}
      \epsfxsize=8.2cm\epsfbox{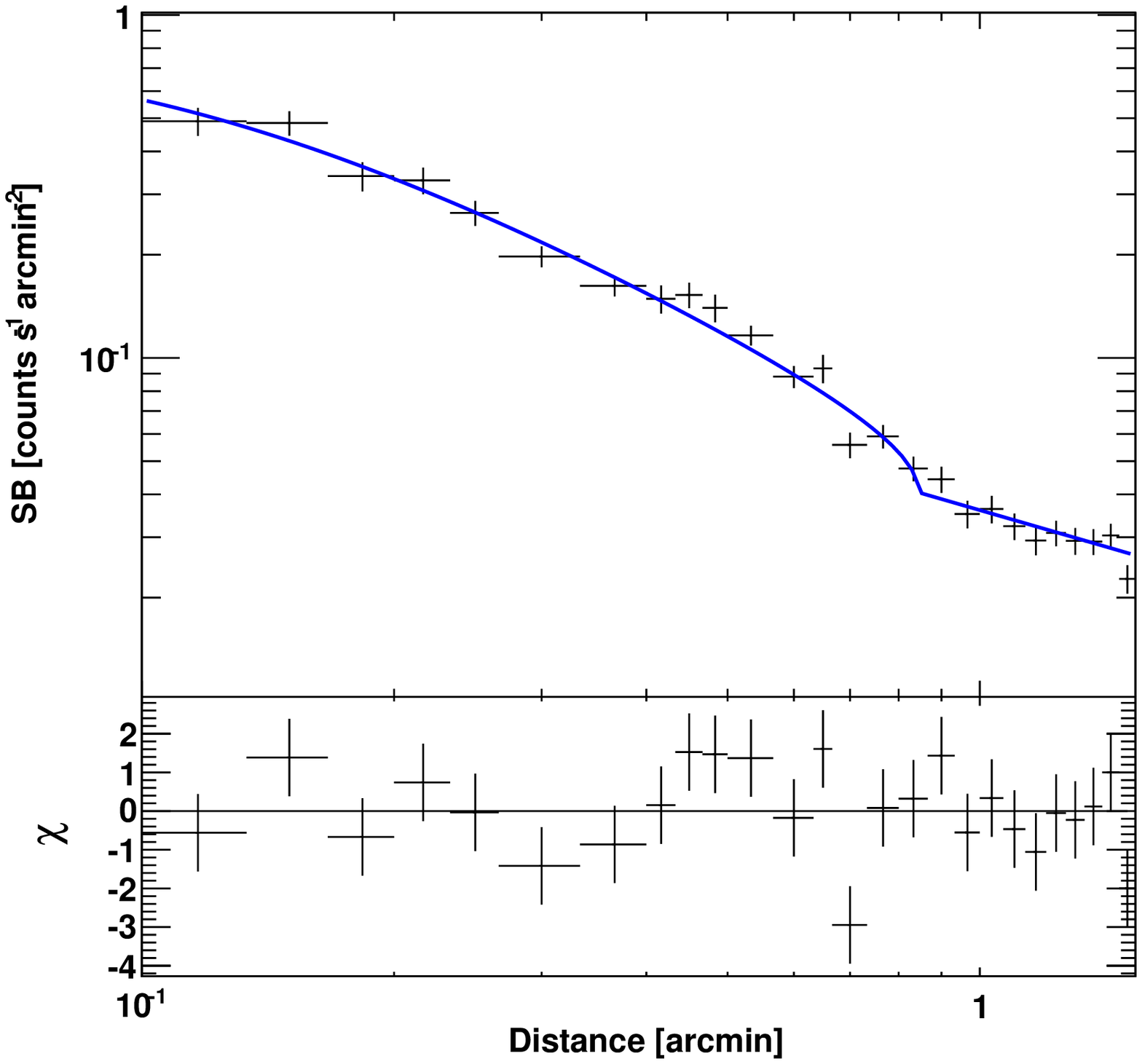}
        \hspace{2cm}
      \epsfxsize=8.2cm\epsfbox{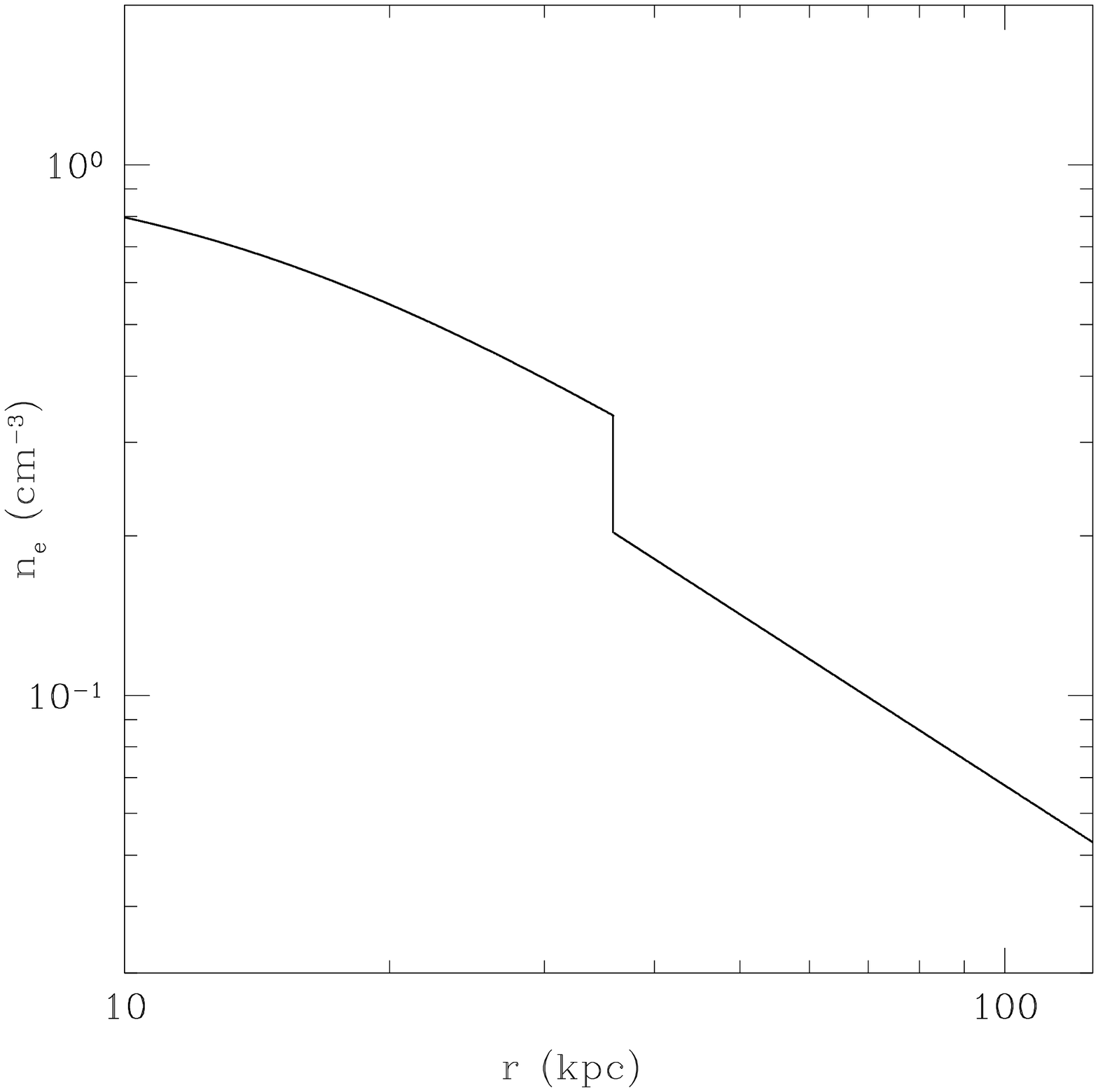}
      \hspace{0.5cm}
      \epsfxsize=8.2cm\epsfbox{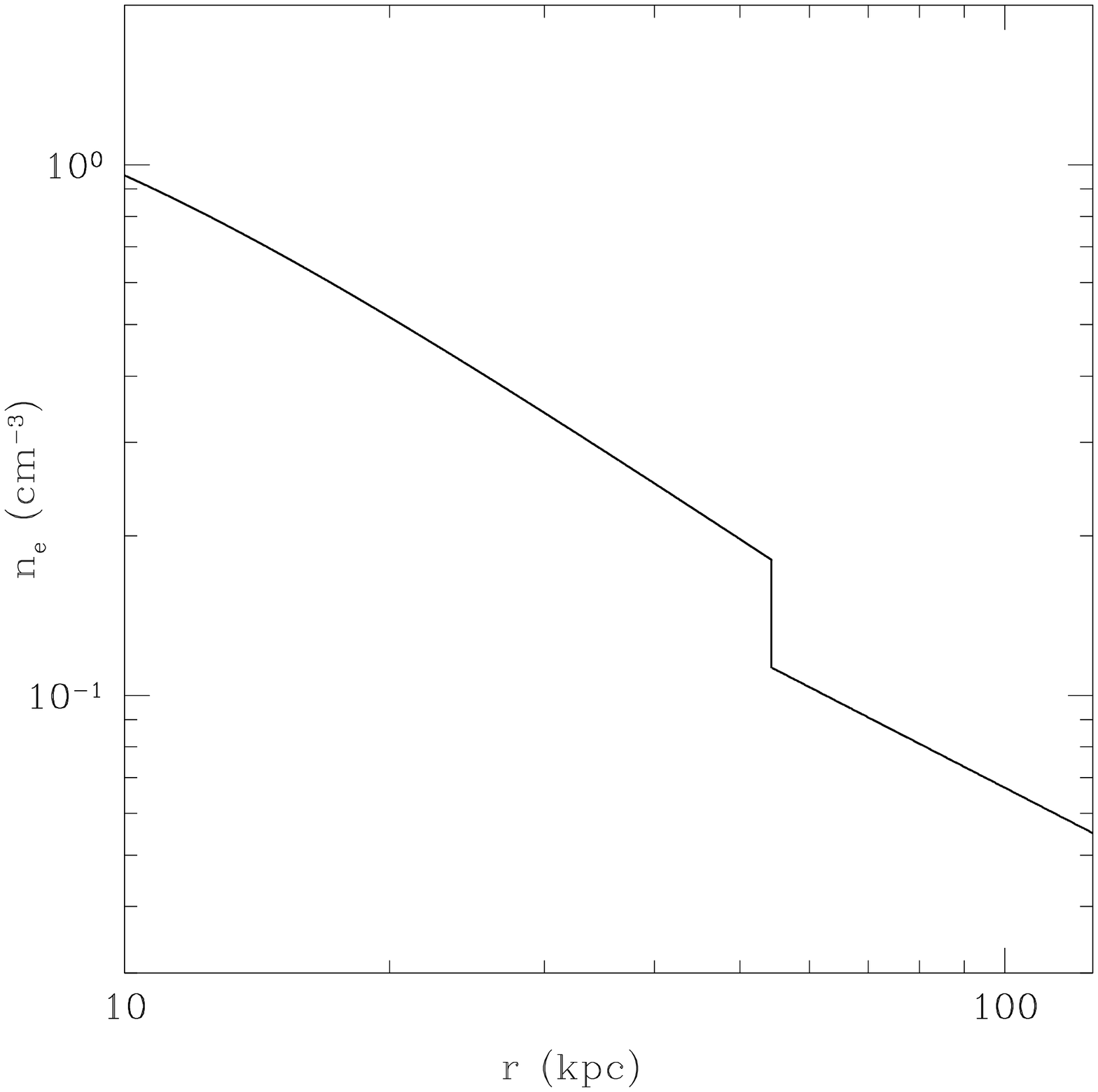}
       \caption{Surface brightness (top panels) and deprojected density (bottom panels) profiles of the central regions of 4C+37.11 for the position angles $0\degr-45\degr$ (left) and $210\degr-255\degr$ (right). Note that position angle $0\degr$ is towards the West. The regions that were used to construct the profiles are depicted in the right panel of Figure \ref{fig:image}. The solid line shows the best fit model to the profiles, which is composed of a $\beta$-model and power law. Note the presence of the jump in surface brightness and density at $34\arcsec$ and $51\arcsec$ for the $0\degr-45\degr$ and $210\degr-255\degr$ wedges, respectively.}
     \label{fig:sb_profiles}
  \end{center}
\end{figure*}

To construct the surface brightness profiles, we extracted the 
brightness in circular wedge regions towards many sectors using
\texttt{PROFFIT}, an interactive software for the analysis of X-ray surface
brightness profiles \citep{2011Eckert}. Two of them
presented surface brightness discontinuities: the northwest (position angles
$0\degr-45\degr$) and southeast ($210\degr-255\degr$) of the cluster
-- the locations of these regions are shown in the right panel of
Figure \ref{fig:image}. The background subtracted $0.5-2$ keV band
profiles are depicted in the top panels in Figure
\ref{fig:sb_profiles}. The profiles demonstrate the presence of
surface brightness jumps, which are located at $\sim 34\arcsec$ and
$\sim 51\arcsec$ central distance on the northwest and southeast,
respectively. To constrain the jump conditions, we fit the surface brightness
profile within each wedge assuming spherical symmetry for the gas density 
and constant gas temperature. The density profile is assumed to follow a $\beta$-model 
and a power law, which is given by the following equations:
\begin{eqnarray}
n(r)=\left\{ \begin{array}{ll}
\renewcommand{\arraystretch}{3}
A\left[1 + (r/r_{\rm c})^2\right]^{-3\beta/2},	
			& \mbox{\hspace{0.9cm} $r \leq r_{\rm cut}$}\\
B \left(r/r_{\rm c}\right)^{-\alpha},	
			& \mbox{\hspace{0.9cm} $r > r_{\rm cut}$}\\
\end{array}
\right.
\label{eq:compression}
\end{eqnarray}
where $r_{\rm{c}}$ is the core radius, and $r_{\rm cut}$ is the radius
where the density abruptly changes. The constants $A$ and $B$ are related by:
\begin{eqnarray}
B = \frac{A \left[1 + (r_{\rm cut}/r_{\rm
      c})^2\right]^{-3\beta/2}}{C \left(r_{\rm cut}/r_{\rm c}\right)^{-\alpha}}
\end{eqnarray}
where $C$ is the density jump. 

The best-fit density profiles are shown in the bottom panels in 
Figure \ref{fig:sb_profiles}. These profiles reveal
density jumps of $1.60_{-0.20}^{+0.29}$ and $1.66_{-0.15}^{+0.15}$ for
the northwest and southeast regions of the cluster, respectively (Table \ref{tab:jump_fit}).
In addition, the position of the jumps are strongly
asymmetric between the two sides of the cluster, since they are at $34
\arcsec$ on the northwest and at $51\arcsec$ on the southeast. 
The best-fit parameters of the fits are listed in Table \ref{tab:jump_fit}.

\begin{table}
\caption{The best-fit jump conditions}
\begin{minipage}{8.75cm}
\renewcommand{\arraystretch}{1.6}
\centering
\begin{tabular}{c c c }
\hline
 & Northwest&  Southeast \\
\hline
$\beta$ & $0.35\pm0.17$ &  $0.37\pm0.05$  \\
$\alpha$  &  $1.07\pm0.05$ & $0.85\pm0.10$  \\
$r_{\rm c}$ [$\arcmin$] &$0.21\pm0.16$  & $0.11\pm0.07$  \\
$r_{\rm{cut}}$ [$\arcsec$] & $34\pm1$  & $51\pm1$ \\
\hline
$n_{\rm 0,cut}/n_{\rm 1,cut}$  &  $1.60^{+0.29}_{-0.20}$ & $1.66\pm0.15$ \\
$T_{\rm 0,cut}/T_{\rm 1,cut}$  &  $0.55\pm0.05$ & $0.46\pm0.08$\\
$p_{\rm 0,cut}/p_{\rm 1,cut}$  &  $0.88\pm0.18$ & $0.76\pm0.15$\\
\hline \\
\end{tabular}
\end{minipage}
\label{tab:jump_fit}
\end{table}

To probe the presence of temperature jumps associated with the 
density jumps, we extract X-ray energy spectra wedges using wedges
with position angles of $0\degr-45\degr$ and $210\degr-255\degr$ for the 
northwest and southeast regions, respectively. The width of the 
regions was $17\arcsec$ for the northwest and $25\arcsec$ for the 
southeast regions, with the boundary positioned at the expected 
location of the jump. The spectra are fit with an optically-thin
thermal plasma emission model (\textsc{apec} in XSPEC). We have fixed
the column density at $N_{\rm H} = 8.2 \times 10^{21} \rm cm^{-2}$ 
and the metal abundances at $A = 0.3$ (Section \ref{sec:temp}). For the northwest 
of the cluster we measure $T_{\rm 0,cut} = 2.64^{+0.12}_{-0.09} $ keV and 
$T_{\rm 1} = 4.84^{+0.36}_{-0.35} $ keV, while for the southeast we obtain 
$T_{\rm 0} = 3.95^{+0.21}_{-0.20}  $ keV and 
$T_{\rm 1} = 8.66^{+1.37}_{-1.15} $ keV. Thus, we obtain a statistically 
significant temperature jump on both sides of the cluster with temperature jumps of
$T_{\rm 0}/T_{\rm 1} = 0.55 \pm{0.05}$ and 
$T_{\rm 0}/T_{\rm 1} = 0.46 \pm{0.08}$ for the northwest and southeast, respectively. 

Based on these data we compare the pressures on both sides of the
surface brightness jumps using $p = n_{\rm e} kT$, where $n_{\rm e}$ 
is the electron density and $kT$ is the gas temperature. 
The derived pressure ratios of $p_0/ p_1 = 0.88\pm0.18$ 
for the jump located on the West, and $p_0 / p_1 =
0.76\pm0.15$ for the  jump at the East, where $p_0$ and $p_1$ 
correspond to the pressures within and beyond the jumps,
respectively.

Thus, the hot gas is in approximate pressure
equilibrium,  implying that the observed sharp surface brightness edges 
most likely originate from the sloshing of the hot gas \citep{2007Markevitch}.

\subsection{Spiral pattern of the hot gas}
The telltale sign of gas sloshing is the presence of a spiral structure in
the hot gas distribution \citep{2007Markevitch}. However, the presence of such underlying
features may not be apparent in the X-ray images due to the strong
surface brightness gradient associated with the cluster. Therefore, we
have produced a ``residual'' X-ray image, in which the merged,
flat-fielded, and background subtracted $0.5-2$ keV band X-ray image
was divided by an azimuthally averaged X-ray image. Note that we
excluded bright point sources and filled their locations with local
background.  To construct the azimuthally averaged image, we described
the surface brightness of the cluster with an elliptical
$\beta$-model, and produced the average image according to the
best-fit properties of this model. Finally, the residual image was
smoothed with a Gaussian with a kernel size of $\sim 2\arcsec$. The  residual ratio 
image removes the strong surface brightness gradients, allowing us to explore faint 
underlying features.

\begin{figure*}[!]
  \begin{center}
    \leavevmode
      \epsfxsize=17cm\epsfbox{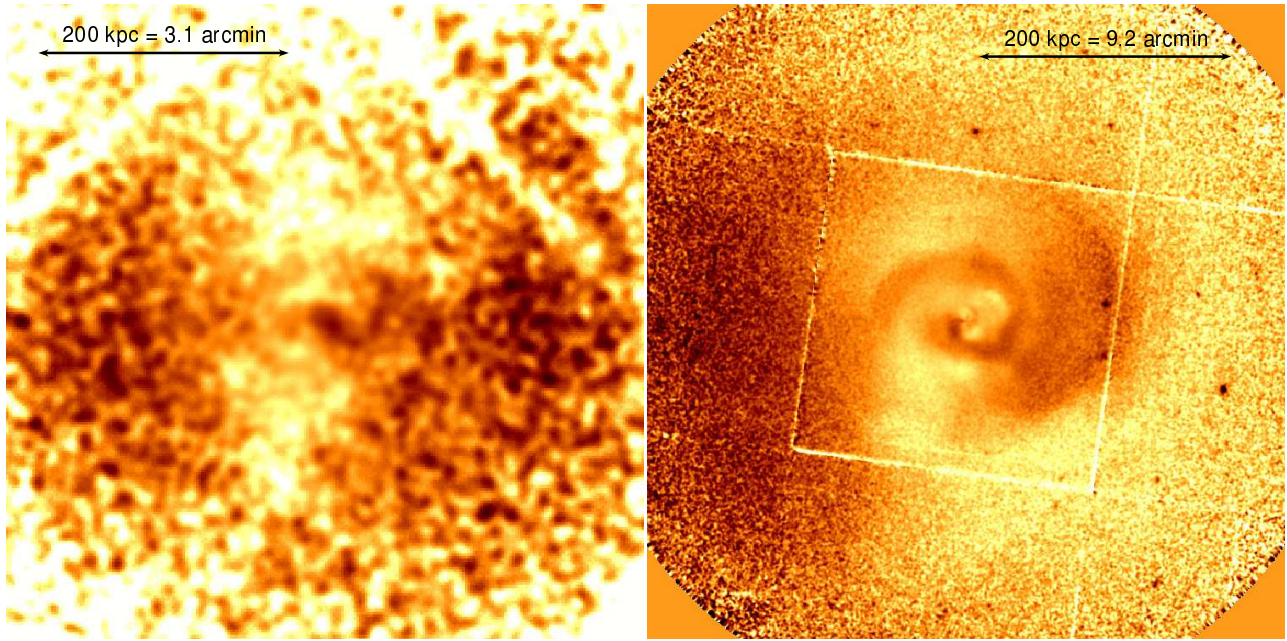}
       \caption{Left: The residual $0.5-2$ keV band \textit{Chandra} image of 4C+37.11 
       showing the underlying structure of the hot gas. To produce this image, we divided 
       the merged X-ray image with an azimuthally averaged image. To enhance the features, 
       the image was smoothed with a Gaussian with a kernel size of 4 pixels. Point sources 
       were excluded and their location were filled with the local emission level. 
       The residual image hints at the presence of a spiral-like feature, 
       which is the characteristic signature of sloshing. Right: The sloshing pattern of 
       Perseus cluster taken from  \citep{2003Churazov}. Note the similar features between the two clusters. 
       Both images depict the same physical regions on the sky. Both clusters 
       show the presence of a knot in the central regions, the excess X-ray emission in the 
       eastern side, and the spiral pattern of the hot gas. Note that the spiral pattern is 
       more graphic in the Perseus cluster, which is due to its proximity and the significantly 
       deeper X-ray observations.}
     \label{fig:perseus_spiral}
  \end{center}
\end{figure*}

The  residual image, shown in the left panel of Figure \ref{fig:perseus_spiral},  hints at the
presence of a spiral pattern in the central regions of the
cluster. Indeed, this feature is very similar to that obtained in
other clusters, such as the Perseus cluster \citep{2003Churazov}.
To directly compare the spiral structure observed in 4C+37.11 and sloshing feature 
observed in Perseus, we compare their images side-by-side in Figure 
\ref{fig:perseus_spiral} at the same physical scale. These images share a number of  similar features, such 
as the bright knot in the center, the excess emission on the eastern side of the 
knot and the fainted emission on the northern side of the spiral pattern. 
Overall, the residual image supports our earlier findings based on the pressure 
profiles, and hints that the hot gas of 4C+37.11 is sloshing. Indeed the overall similarity of
scale and structure to that seen in Perseus suggests a similar time since the last merger event.

\section{Small scale structure}
\label{sec:small}
In galaxies, galaxy groups, and galaxy clusters, cavities in the hot gas distribution 
are widely observed, and hint at the presence of AGN outbursts. Indeed, powerful radio 
sources are capable of drastically increasing the entropy of the hot
gas, inflating the gas distribution and reducing its density. 

\begin{figure*}[!]
  \begin{center}
    \leavevmode
      \epsfxsize=10.6cm\epsfbox{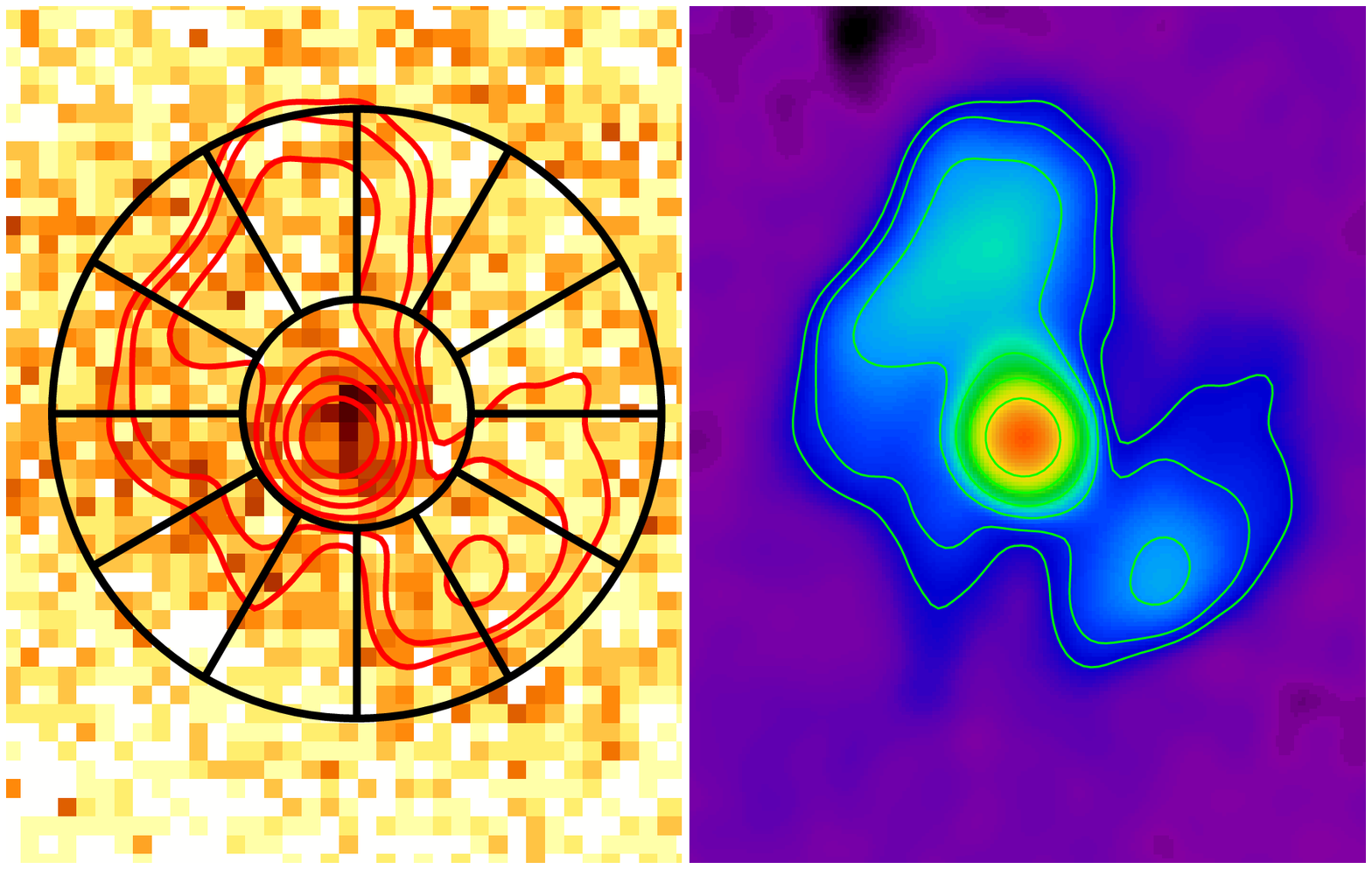}    
      \hspace{0.65cm}
      \epsfxsize=6.5cm\epsfbox{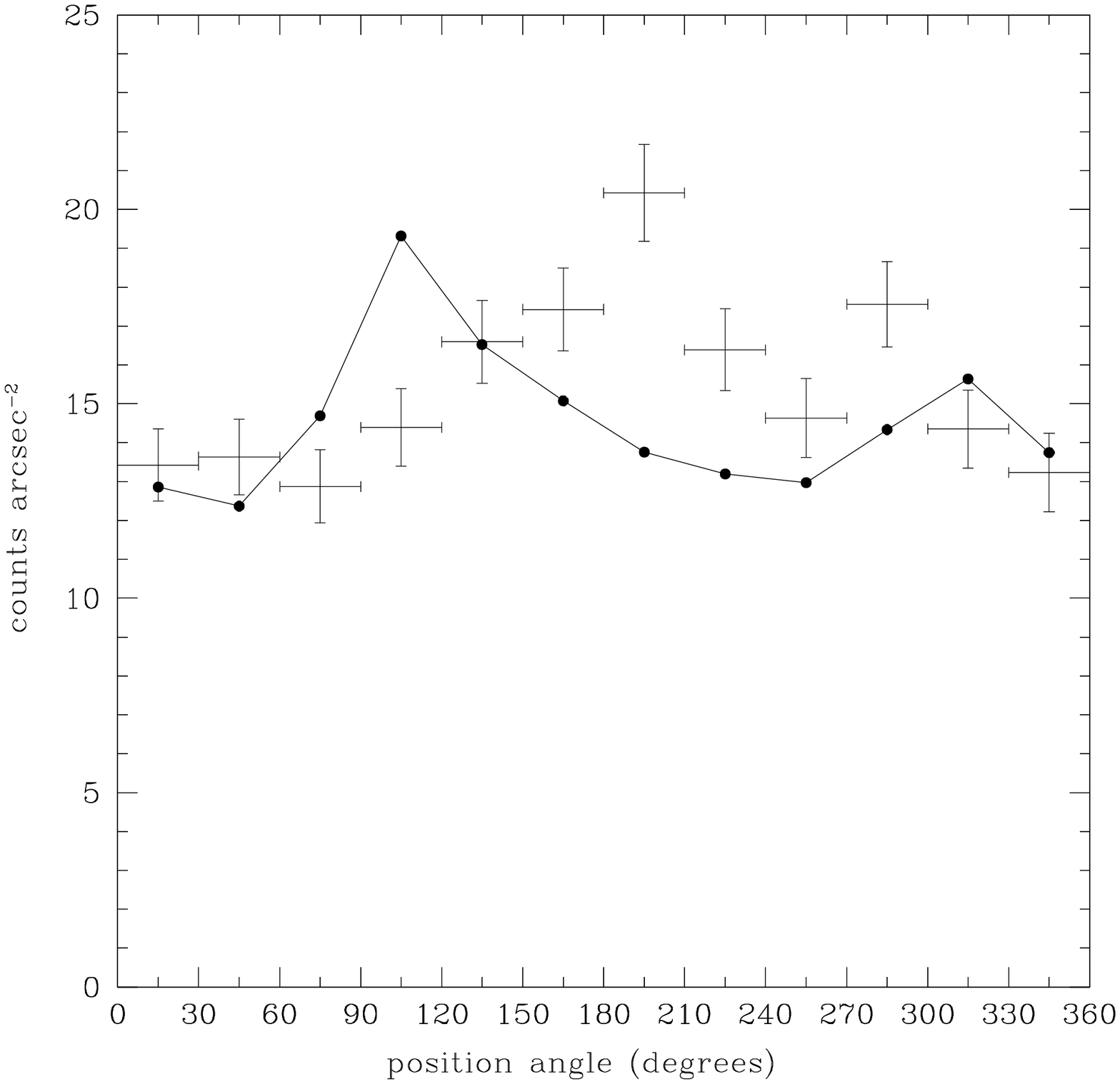}
       \caption{Left: $0.5-2$ keV band merged \textit{Chandra} X-ray
         image of the central $17\arcsec \times 22\arcsec$ region of
         4C+37.11. Overlayed are the intensity levels of the $1.4$ GHz VLA observations. Note that the intensity levels are spaced logarithmically. There is no clear correlation between the decrements in the X-ray surface brightness distribution and the position of the radio lobes. Right: X-ray and radio surface brightness profiles extracted from circular wedges with $3\arcsec-8\arcsec$ radii. Points with error bars represent the X-ray surface brightness, while the connected points show the radio intensity levels, which were arbitrarily normalized to match the X-ray surface brightness level. The position angle $0\degr$ is toward the West. Note the absence of an anti-correlation between the X-ray and radio intensity levels.}
     \label{fig:radio}
  \end{center}
\end{figure*}

Radio observations of 4C+37.11 point at the existence of radio lobes
in the central $20\arcsec$ region of the host elliptical. In Figure
\ref{fig:radio}, we show the $0.5-2$ keV band merged \textit{Chandra}
X-ray image of the cluster along with the intensity levels of the 1.4
GHz VLA data. Interestingly, the X-ray image of 4C+37.11 (see Figure \ref{fig:radio}) hints 
that at the position of the radio lobes a decrement in the X-ray surface brightness 
may be present. While the  distribution of the hot X-ray gas exhibits
strong asymmetries, the decrements in the X-ray surface brightness are
not exactly associated with the position of the radio lobes. 
To quantitatively probe the possible anti-correlation between the
X-ray and radio intensity levels, 
we extract the $0.5-2$ keV band X-ray and $1.4$ GHz radio surface
brightness profiles in annular wedges.
The profile, shown in the right panel of Figure \ref{fig:radio}, shows that there is
no clear anti-correlation between the X-ray and radio surface
brightness, which is confirmed by a two sample Kolmogorov-Smirnov test between the X-ray and reciprocal radio
fluxes that excludes similarity at $p = 2.05 \times 10^{-5}$.
However, a weak anti-correlation might be suppressed by projection effects.

\section{Discussion}

The detection of likely sloshing features in the X-ray halo of 4C+37.11
allows us to place constraints on the interaction which created these structures.
While the presence of the radio outflow associated with the CSO raises the
possibility that AGN-driven bubbles can affect the inner halo structure, the
presence of edges at large radius and the overall spiral appearance of the
residual image (Figure \ref{fig:perseus_spiral}) makes it likely that a dynamical interaction has
defined the basic disturbance. Further, given the marked similarity to the Perseus
halo, it is productive to compare with the simulations of  \citet{2006Ascasibar} and 
\citet{2010ZuHone}. These simulations assume a 1:5 merger mass ratio and
impact parameter b = 500 kpc and result in spiral gas distribution matching
that seen in Perseus at times 1$-$2 Gyrs after the first passage of the interacting
sub-cluster. At this time this sub-cluster makes its second approach, but the
cores have not yet merged. The stellar component of such a sub-cluster for 4C+37.11
is not obvious; it might be identified with the nearest, 2.5 mag fainter,
galaxy about 25$^\prime$ ($\sim27$ kpc) projected distance away or with a more
distant object, but kinematic studies are needed to test such association.

We next consider the implications for the unique tight double radio
nucleus of this source. Recalling that VLBI studies of over 3000 AGN have
found no other resolved $<10$ pc-scale system \citep{2011Burke} , 
it is important to consider whether other unusual system properties (such as the bright X-ray halo) can
inform us about the nucleus. The rarity of resolved double nuclei implies a short
$<0.5$ Gyr time between galaxy core merger and GR-driven inspiral \citep{2011Burke}.
In turn this implies that some source other than dynamical friction drives the
BH cores to radii where GR can take over. So we must ask: why is 4C+37.11
seen at 7pc projected separation?

Broadly speaking there are two possibilities. Perhaps we are simply catching
this binary shortly after core merger when the black holes are still approaching
each other. Alternatively, the typically dominant merger mechanism (likely mediated
by interactions with dissipative circumbinary gas) is under-performing in
4C+37.11, leaving it stalled at resolvable scale. At face value the presence of
large scale sloshing structures eliminates the first possibility. Here we adopt
the $\sim 1-2$ Gyrs time scale since the last major interaction implied by the
spiral morphology and we note that the cores of the galaxies in this
interaction should remain unmerged.  Thus the interaction of the two cores
contributing the observed, resolved supermassive black hole binary must have
occurred even earlier, implying that these holes have remained unmerged for several Gyr.
This is substantially longer than the bounds on typical merger timescale.

We see that our X-ray halo studies, while not directly probing at
the black hole binary scale, have allowed us to conclude that unusual core
properties must exist in 4C+37.11, permitting the black hole
to languish at $<10$ pc separation for several Gyrs. Dynamical studies
of the core, and especially studies of the atomic and molecular gas
will be helpful in directly exploring conditions in the circumbinary environment.


\section{Conclusions}

We present results from {\em Chandra} observations of the
cluster-scale X-ray halo surrounding 4C+37.11, a nearby ($z$=0.055) galaxy that hosts 
the closest known resolved supermassive black hole binary. We derive
within $r_{500}$ a total mass of $M_{500} = (2.5 \pm 0.2) \times 10^{14} \rm ~
M_\odot$, a gas mass of $M_{\rm g,500} = (2.2 \pm 0.1) \times 10^{13} \rm ~
M_\odot$ an X-ray bolometric luminosity of $L_{\rm bol,500} = (1.10
\pm 0.01) \times
10^{44} \rm ~ erg~s^{-1}$, and a temperature of $kT = 4.6 \pm 0.2
~ \rm keV$. The gas mass fraction within $r_{500}$ is $f_{\rm g} =
0.09 \pm 0.01$, in
agreement with the expected value, given the cluster temperature. We present
total mass, gas mass, and entropy profiles. Evidence of gas sloshing comes from
extraction of surface brightness and temperature profiles in selected wedges, where we
show that despite the density jump, the pressure is continuous along
the contact discontinuities. We conclude that the 
host of the radio galaxy 4C+37.11 is probably a relaxed fossil cool-core
cluster that has been mildly disturbed as a smaller group or cluster 
interacted gravitationally during close approach and ``sloshed'' the cool gas residing at its center. The interaction driving the sloshing
appears to have occurred $1-2$ Gyrs ago, while the interaction producing the binary
black hole nucleus occurred even earlier. This allows us to conclude that
the supermassive black hole binary in 4C+37.11 is not young and has stalled
outside the present orbital separation for several Gyrs, longer than the time
scale inferred for typical black hole binary coalescence. 

\acknowledgments

\noindent
The authors thank Eugene Churazov for kindly providing the Perseus
residual image, Maxim Markevitch for helpful discussions, and Alexey
Vikhlinin and Georgiana Ogrean for providing software. 
F.A-S. acknowledges support from {\em Chandra} grant G03-14131X. 
\'A.B., W.R.F., C.J. are supported by the Smithsonian Institution. 
R.W.R was supported in part by NASA grant G04-15116A.

\end{document}